\documentclass[12pt]{article}
\usepackage{a4wide,epsf,epsfig}

\newcommand{\gsim}{
\raisebox{-4pt}{$\,\stackrel{\textstyle >}{\sim}\,$}}

\begin{document}

\begin{flushright}
WU B 00-15 \\
hep-ph/0007277
\end{flushright}

\begin{center}
\vskip 3.5\baselineskip
{\Large\bf Skewed parton distributions and the scale dependence of the 
           transverse size parameter}
\vskip 2.5\baselineskip

C. Vogt

\vskip \baselineskip

 Fachbereich Physik, Universit\"at Wuppertal, 42097 Wuppertal,
         Germany 
\vskip \baselineskip

\vskip 2.5\baselineskip

\textbf{Abstract} \\[0.5\baselineskip]

\parbox{0.9\textwidth}{We discuss the scale dependence of a skewed parton
distribution of the pion obtained from a generalized light-cone wave function 
overlap formula. Using a simple ansatz for the transverse momentum dependence 
of the light-cone wave function and restricting ourselves to the case of a zero
skewedness parameter the skewed parton distribution can be expressed through 
an ordinary parton distribution multiplied by an exponential function. Matching
the generalized and ordinary DGLAP evolution equations of the skewed and 
ordinary parton distributions, respectively, we derive a constraint for the 
scale dependence of the transverse size parameter which describes the width of
the pion wave function in transverse momentum space. This constraint has 
implications for the Fock state probability and valence quark distribution. 
We apply our results to the pion form factor.}

\vskip 1.5\baselineskip
\end{center}

Skewed parton distributions (SPDs) provide a link between exclusive and 
inclusive quantities of QCD \cite{Mueller,Ji,Radyushkin}. Among some of their 
well known properties is their relation to ordinary parton distributions
and hadronic form factors via so-called reduction formulas. Moreover, their
evolution behaviour has been investigated and generalized DGLAP evolution 
equations have been derived. Only few is known, however, about their 
particular form and so for applications to physical processes one has to resort
to specific models. The authors of \cite{DFJK} discussed SPDs in the context
of soft contributions to large angle Compton scattering and form factors and
proposed a generalized Drell-Yan overlap formula \cite{DrellYan} for SPDs in 
terms of light-cone wave functions (LCWFs). It was shown that in a special 
kinematical region, where the plus component of the momentum transfer and
consequently the skewedness parameter vanishes, and with a Gaussian ansatz for
the transverse momentum dependence of the LCWFs, SPDs can be expressed by 
ordinary parton distribution functions multiplied by an exponential $t$ 
dependence. 

In the present paper we will investigate the consequences of the evolution
equations for this phenomenological model of SPDs. As we will argue below, the 
combined evolution of the SPD and the ordinary parton distribution enforces a 
condition upon the scale dependence of the transverse size parameter. 
This parameter appears in the SPD through the transverse momentum dependence of
the LCWFs and describes the width of the wave function in $k_\perp$-space. 
We will derive a model dependent intregro-differential equation for the scale 
dependence of 
the transverse size parameter, which we will solve numerically. Moreover, we 
will show that the scale dependent transverse size parameter induces a scale 
dependence of the Fock state probability of the lowest Fock state and the 
corresponding valence distribution. As an application to a physical process it
is natural to consider soft overlap contributions to the pion form factor. We 
conclude with our summary.

In this work we use the conventions of Radyushkin \cite{Radyushkin} and denote
the SPD of a parton with flavour $a$ in the pion by 
$\widetilde{{\cal F}}^a_\zeta(x,t)$. It is defined by a bilocal matrix element
of quark field operators:
\begin{equation}
 p^+\int\frac{dz^-}{2\pi}\,e^{ixp^+z^-} \langle\pi(p')|\overline{\psi}_a(0)
 \gamma^+\psi_a(z^-)|\pi(p)\rangle=\widetilde{\cal F}^{a}_\zeta(x,t)\,(p+p')^+,
\end{equation}
where the notation $\psi_a(z^-)$ indicates that the argument of the operator
has vanishing light-cone plus and transverse components. The following 
reduction formulas \cite{Ji,Radyushkin} are general properties of SPDs. In the
forward case one regains ordinary parton distributions:
\begin{equation}
 \widetilde{{\cal F}}^a_{\zeta=0}(x,t=0)=q_a(x),
\end{equation}
and by integrating the valence distribution one obtains the pion form factor:
\begin{equation}
 F_\pi(t)=\int_0^1 dx\,\widetilde{\cal F}^{v}_\zeta(x,t), \label{pionff}
\end{equation}
where we have defined $\widetilde{\cal F}^v_\zeta(x,t)=
\widetilde{\cal F}^a_\zeta(x,t)-\widetilde{\cal F}^{\bar{a}}_\zeta (x,t)$ 
and used $e_u-e_d=1$.

Our starting point is the overlap formula for SPDs in the region 
$\zeta<x<1$ \cite{DFJK}:
\begin{equation}
 \widetilde{{\cal F}}^{a(N)}_\zeta(x,t)=\frac{(1-\zeta)^{1-N/2}}{1-\zeta/2}
     \sum_{l,\beta}\int[dx]_N [d^2{\bf k}_\perp]_N \, \delta(x-x_l) 
          \Psi_{N\beta}^*(x_i',{\bf k}'_{\perp i}) \, 
          \Psi_{N\beta}(x_i,{\bf k}_{\perp i}),  \label{spd overlap}
\end{equation}
where $N$ denotes a particular Fock state, the summation index $l$ runs over
all quarks of flavour $a$ and $\beta$ refers to different spin flavour 
combinations of partons in a given $N$-particle Fock state.
The integration measures are defined by
\begin{eqnarray}
 &&[dx]_N\equiv\prod_{n=1}^N dx_n \,\delta\bigg(1-\sum_m x_m\bigg), \nonumber\\
 &&[d^2{\bf k}_\perp]_N\equiv\frac{1}{(16\pi^3)^{(N-1)}}\prod_{n=1}^N
    d^2{\bf k}_{\perp n} \, \delta^{(2)}\bigg(\sum_m {\bf k}_{\perp m}\bigg) 
\end{eqnarray}
and the arguments of the initial and final wave function are related by
\begin{eqnarray}
 x_i'&=&\frac{x_i}{1-\zeta} \quad,\quad \;\,{\bf k}'_{\perp i}
    ={\bf k}_{\perp i}-\frac{x_i}{1-\zeta}{\bf \Delta}_\perp, \nonumber\\
 x_j'&=&\frac{x_j-\zeta}{1-\zeta} \;\; , \;\quad {\bf k}'_{\perp j}
    ={\bf k}_{\perp j}+\frac{1-x_j}{1-\zeta}{\bf \Delta}_\perp, 
\end{eqnarray}
with $j$ being the index of the active parton and $i$ being the index of the 
spectator partons. ${\bf \Delta}_\perp$ is the transverse momentum transfer
between the initial and final hadron.

Following the authors of \cite{DFJK,JakobKroll,BolzKroll}, we write the soft 
$N$-particle LCWF $\Psi_{N\beta}$ of the pion in terms of the distribution 
amplitude $\phi_{N\beta}$ and a transverse momentum dependent part, for which 
we make a Gaussian ansatz: 
\begin{equation}
 \Psi_{N\beta}(x_i,{\bf k}_{\perp i})={\cal N}_{N\beta} \phi_{N\beta}(x_i)
  \frac{(16 \pi^2 a_N^2)^{N-1}}{x_1 x_2\dots x_N} \exp\bigg[
     -a_N^2\sum_{i=1}^N\frac{{\bf k}^2_{\perp i}}{x_i}\bigg], \label{lcwf}
\end{equation}
where ${\cal N}_{N\beta}$ is a normalization constant and, for obvious reasons,
$a_N$ is called the transverse size parameter of the $N$-particle Fock state. 

As we have already mentioned we will restrict ourselves to the special case of
a zero skewedness parameter, i.e. $\zeta=0$. It has been shown in \cite{DFJK}
that the Gaussian ${\bf k}_\perp$-dependence then leads to the following simple
representation of an SPD in terms of an ordinary parton distribution function
multiplied by an exponential:
\begin{equation}
 \widetilde{{\cal F}}^{a(N)}_{\zeta=0}(x,t)=q_a^{(N)}(x)\,
    \exp\bigg[\frac{1}{2}a_N^2\frac{\bar{x}}{x} t\bigg], \label{spd N}
\end{equation}  
where we use the common notation $\bar{x}\equiv 1-x$. The origin of the 
appearance of the parton distribution in (\ref{spd N}) is its relation to 
LCWFs via the expression~\cite{BHL} 
\begin{equation}
 q_a^{(N)}(x)=\sum_{l,\beta} \int[dx]_N [d^2{\bf k}_\perp]_N \, 
      \delta(x-x_l)|\Psi_{N\beta}(x_i,{\bf k}_{\perp i})|^2. \label{pdf}
\end{equation}  
As will become obvious immediately it is useful to approximate the transverse
size parameters of all Fock states by a common value $a_\pi$, i.e. we set
\begin{equation}
 a_\pi\simeq a_N\quad {\rm\; for\; all\; } N. \label{approx api}
\end{equation}
This is in general a rather rough approximation. However, from the exponential
in Eq.~(\ref{spd N}) together with the fact that with increasing $N$ the 
functions $q_a^{(N)}(x)$ are proportional to increasing powers of 
$(1-x)$~\cite{DFJK} it is clear that large $x$ dominate at large~$t$ and only 
a few of the lowest Fock states contribute to phenomenological applications.
We can now sum over all Fock states,
\begin{equation}
 \widetilde{{\cal F}}^{a}_{\zeta=0}(x,t)=\sum_N 
       \widetilde{{\cal F}}^{a(N)}_{\zeta=0}(x,t),
\end{equation}
so as to obtain a representation in terms of the full quark distribution 
function. Since we will discuss the pion form factor later on and in order to 
avoid complications in our discussion of evolution by quark-gluon mixing we 
consider the valence distribution from now on: 
\begin{equation}
 \widetilde{{\cal F}}^{v}_{\zeta=0}(x,t)=q_v(x)\,\exp\bigg[\frac{1}{2}a_\pi^2
   \frac{\bar{x}}{x} t\bigg]. \label{valence spd}
\end{equation}  
The SPD (\ref{valence spd}) is completely independent of the particular form of
the pion distribution amplitude. Thus, we need not to specify the pion 
distribution amplitudes $\phi_{N\beta}$ in (\ref{lcwf}). A corresponding 
expression for the SPD of the nucleon has been suggested in 
Refs.~\cite{DFJK,Radyushkin2} and in case of the pion in Ref.~\cite{Afanasev}
and also recently by the authors of~\cite{BakulevRuskov}.

At this stage, the SPD (\ref{valence spd}) depends on a scale $\mu^2$ only 
through the ordinary parton distribution. However, as we are going to show the
evolution equation of the SPD forces the transverse size parameter $a_\pi$ to 
be scale dependent as well.
The scale dependence of the l.h.s. of expression (\ref{valence spd}) is 
described in terms of a generalized DGLAP evolution equation, where for 
$\zeta=0$ the modified evolution kernels are known to reduce to the ordinary 
DGLAP kernels \cite{Ji,Radyushkin}:
\begin{equation}
 \mu^2\frac{\partial}{\partial\mu^2}\,\widetilde{\cal F}^v_{\zeta=0}(x,t;\mu^2)
  =\frac{\alpha_s(\mu^2)}{2\pi}\,\int_x^1 \frac{dy}{y} \, P_{qq}
 \Big(\frac{x}{y}\Big) \,\widetilde{\cal F}^v_{\zeta=0}(y,t;\mu^2). \label{lhs}
\end{equation}   
The r.h.s. of (\ref{valence spd}) is given in terms 
of a parton distribution obeying ordinary DGLAP evolution, multiplied by an 
exponential. Obviously, both the evolution equations of the SPD and of the 
parton distribution can only be fulfilled simultaneously if 
\begin{equation}
 a_\pi=a_\pi(\mu^2),
\end{equation}
so that for the r.h.s. of (\ref{valence spd}) we have:
\begin{eqnarray}
 &&\mu^2\frac{\partial}{\partial\mu^2}\,\Bigg\{q_v(x;\mu^2)\,\exp\bigg[
  \frac{1}{2} a_\pi^2(\mu^2)\frac{\bar{x}}{x}\,t\bigg]\Bigg\} \nonumber\\
    &&=\exp\bigg[\frac{1}{2} a_\pi^2(\mu^2)\frac{\bar{x}}{x}\,t\bigg]\,
       \Bigg\{\frac{\alpha_s(\mu^2)}{2\pi}\,\int_x^1 \frac{dy}{y}\, 
       P_{qq}\Big(\frac{x}{y}\Big) \,q_v(y;\mu^2)
      +q_v(x;\mu^2)\,\frac{1}{2}\frac{\bar{x}}{x}\,t\,\mu^2\,
        \frac{d a_\pi^2(\mu^2)}{d \mu^2}\Bigg\}. \label{rhs} \nonumber\\
\end{eqnarray}
Equating expressions (\ref{lhs}) and (\ref{rhs}) and resolving for the
derivative of $a_\pi(\mu^2)$ we find an integro-differential equation for
the scale dependence of the transverse size parameter: 
\begin{eqnarray}
 \mu^2\;\frac{d a_\pi^2(\mu^2)}{d \mu^2}=&-&\bigg[\frac{1}{2}
 \frac{\bar{x}}{x}\,t\,q_v(x;\mu^2)\bigg]^{-1} \nonumber \\  &\times&
 \frac{\alpha_s(\mu^2)}{2\pi}\,\int_x^1 \frac{dy}{y}\,P_{qq}\Big(\frac{x}{y}
 \Big)\Bigg\{1-\exp\bigg[\frac{1}{2}a_\pi^2(\mu^2)\Big(\frac{\bar{y}}{y}-
 \frac{\bar{x}}{x}\Big)\,t\bigg]\Bigg\}\,q_v(y;\mu^2). \label{api dgl}
  \nonumber\\
\end{eqnarray}
This equation is a consequence of our particular model~(\ref{valence spd}).
As we can see immediately, the evolution of the transverse size parameter is
driven by the difference of the SPD and the ordinary parton distribution. This 
equation may be solved numerically using an iterative method similar to the
one employed for the numerical solution of ordinary DGLAP equations, see for
instance \cite{Miyama}. In order to determine the initial value of $a_\pi$
we consider the LCWF of the lowest Fock state. It is commonly accepted that the
form of the pion's two-particle distribution amplitude is close to the 
asymptotic one \cite{Musatov}, $\phi^{\rm as}_2(x)=6\,x\bar{x}$, to which we 
will restrict ourselves in the following. The parameters of the two-particle
LCWF are then completely fixed from various decay processes \cite{BHL}. The 
normalization follows from the $\pi^+\to\mu^+\nu_\mu$ decay and it is given by
${\cal N}_2=f_\pi/(2\sqrt{6})$, where $f_\pi\simeq 131$~MeV is the well known 
pion decay constant. The two-photon decay of the uncharged pion, 
$\pi^0\to\gamma\gamma$, leads to a constraint for the transverse size 
parameter, for which a value of $a_\pi=0.86$~GeV$^{-1}$ is found.
As the corresponding scale we choose $\mu_0^2=0.25$~GeV$^2$ which is a typical
scale for light mesons~\cite{CZ}. From now on we will employ the GRS 
parametrization~\cite{GRS} of the valence quark distribution $u_v(x;\mu^2)$. 

It is interesting to note that the singularity of the splitting function is 
canceled by the expression in curly brackets of Eq.~(\ref{api dgl}). As we can 
see further, there are two free variables in Eq.~(\ref{api dgl}): the 
longitudinal momentum fraction $x$ and the momentum transfer $t$. This means 
that our approach does not a priori exclude an explicit $x$- and $t$-dependence
of $a_\pi$. Intuitively, however, we expect transverse quantities such as
the mean square transverse momentum $\langle {\bf k}_\perp^2\rangle$  neither
to depend on the longitudinal momentum nor on the momentum transfer. 
The numerical investigation indeed only shows a very mild variation of 
$a_\pi(\mu^2)$ at different values of large $x$ and $t$. If this were not the 
case the ansatz (\ref{valence spd}) would have to be abandoned. For 
definiteness, in what follows we will choose $x$ in Eq.~(\ref{api dgl}) to be 
the mean value of the momentum fraction contributing to the pion form factor 
which is given by 
\begin{equation}
 \langle x\rangle_t:=\frac{1}{F_\pi(t)}\int_0^1 dx\;x\;
        \widetilde{\cal F}^v_{\zeta=0}(x,t). \label{average x}
\end{equation}
Taking $-t=10$~GeV$^2$ we find a value of $\langle x \rangle_t=0.75$. Varying 
$-t$ between 1~GeV$^2$ and 10~GeV$^2$ $a_\pi(\mu^2)$ only changes by less than
4\%.
\begin{figure}[t]
\begin{center}
\psfig{file=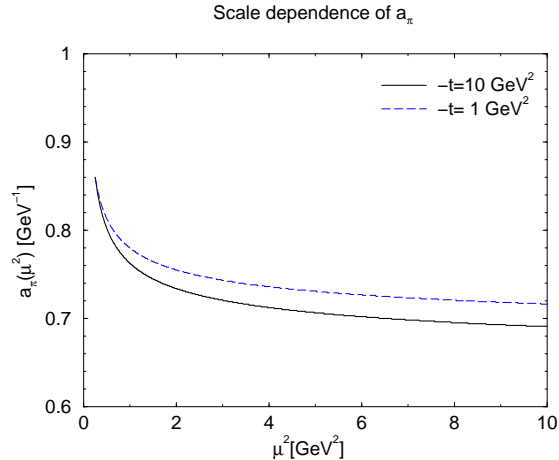,bb=45 70 560 675,width=6cm,angle=-90}
\caption{Scale dependence of the transverse size parameter $a_\pi$ 
         with an initial value of $a_\pi=0.86$ GeV$^{-1}$ at 
         $\mu_0^2=0.25$ GeV$^2$. As described in the text we have chosen 
         $x=\langle x \rangle_t=0.75$. The two curves demonstrate the weak
         dependence on $t$.} \label{plot api}
\end{center}
\end{figure}
\begin{figure}[t]
\begin{center}
\psfig{file=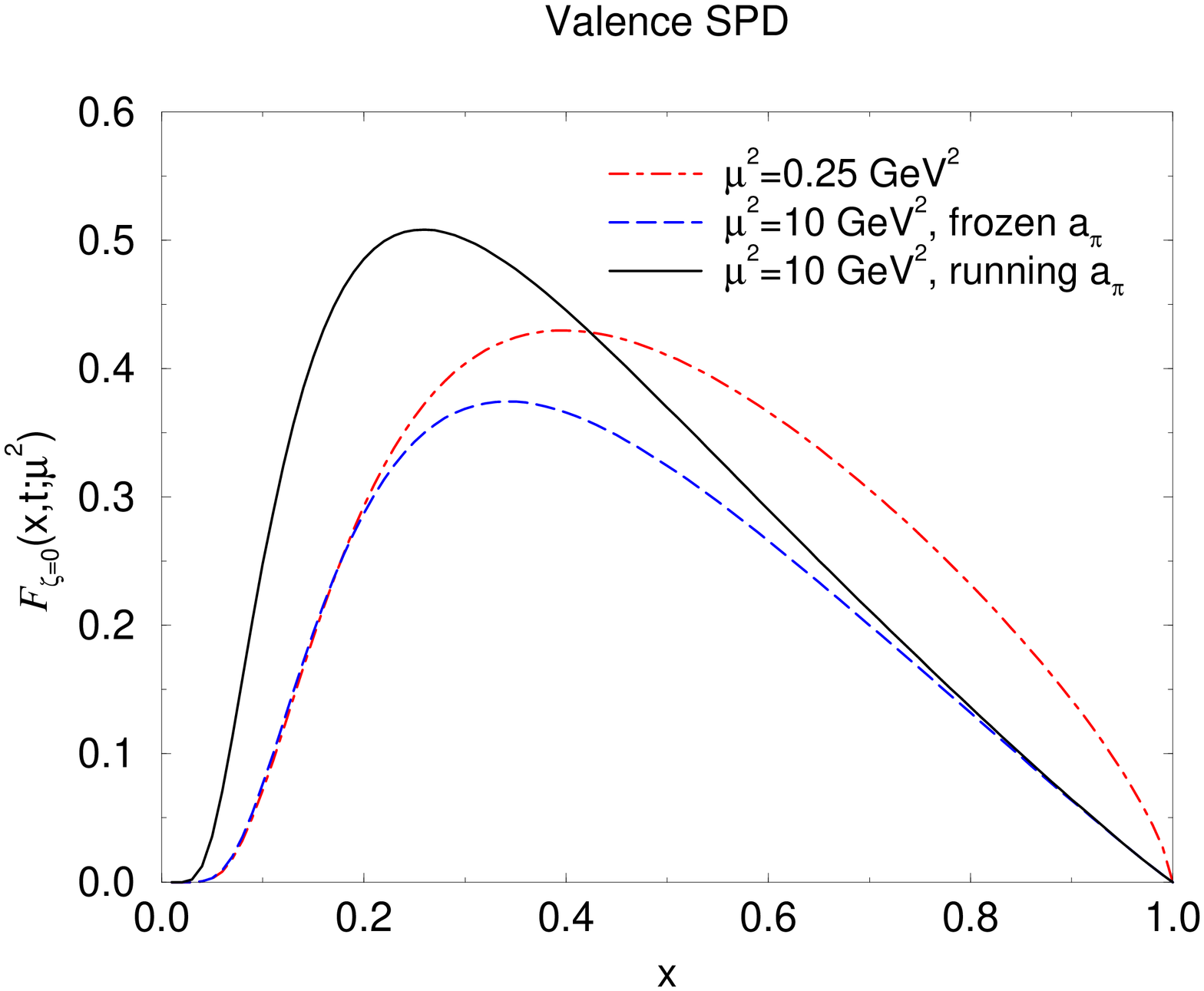,bb=45 70 560 675,width=6cm,angle=-90}
\caption{Evolution of the valence SPD (\ref{valence spd}) with constant and 
         running $a_\pi$. We use the GRS parametrization of the forward valence
         distribution as stated in the text. At the starting scale 
         $\mu_0^2=0.25$ GeV$^2$ both curves naturally coincide.
         The momentum transfer $-t$ is arbitrarily set to 1 GeV$^2$.}  
      \label{plot spd}
\end{center}
\end{figure}

The scale dependence of the transverse size parameter is shown in 
Fig. \ref{plot api}. As can be expected $a_\pi$ depends only moderately on 
$\mu^2$ and the decrease is weakened with increasing scale. 
In Fig. \ref{plot spd} we have plotted the valence
SPD (\ref{valence spd}) at two different scales. In order to see which 
quantitative effect the scale dependence of $a_\pi$ has on the SPD, we compare
$\widetilde{\cal F}^v_{\zeta=0}(x,t;\mu^2)$ with frozen (dashed line) and 
running $a_\pi$ (solid line), respectively, at $\mu^2=10$~GeV$^2$, where the 
scale dependence of the transverse size parameter causes a shift of the SPD 
towards smaller momentum fractions. At $\mu_0^2=0.25$~GeV$^2$ both curves 
coincide by definition (dot-dashed line). 

At constant $a_\pi$ the parton distribution of the lowest Fock state which 
results from the LCWF (\ref{lcwf}) is scale independent. With increasing scale,
however, one expects a damping of the parton distributions of lower Fock 
states since an increasing number of virtual quark-antiquark pairs and gluons 
is produced so that higher and higher Fock states become occupied. 
Quantitatively, this damping effect emerges through the Fock state probability
in our approach. The Fock state probability of the $N$-th Fock state is defined
by 
\begin{equation}
 P_N\equiv \sum_\beta\int [dx]_N [d^2{\bf k}_\perp]_N 
       |\Psi_{N\beta}(x_i,{\bf k}_{\perp i})|^2.
\end{equation}
The running $a_\pi$ induces a scale dependent Fock state probability. As we 
have already discussed below Eq.~(\ref{api dgl}) the normalization of the LCWF
of the pion's lowest Fock state is fixed so that for $P_2$ we have explicitly 
\begin{equation}
 P_2(\mu^2)=2 \pi^2\, f_\pi^2\, a_\pi^2(\mu^2),
\end{equation} 
which coincides with the well known value of $P_2=0.25$ at 
$\mu_0^2=0.25$ GeV$^2$ due to our choice of the initial value of $a_\pi$. 
As we can see immediately the Fock state probability of the lowest Fock state
decreases with increasing scale since it is proportional to $a_\pi^2(\mu^2)$.
The value of $P_2$ at $\mu^2=100$ GeV$^2$ is reduced to about 0.14.

\begin{figure}[t]
\begin{center}
\psfig{file=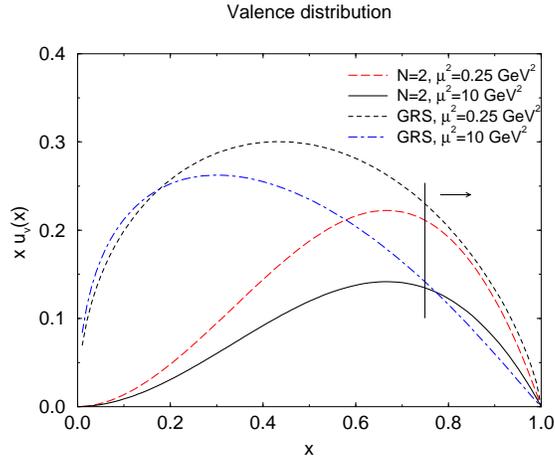,bb=45 70 560 675,width=6cm,angle=-90}
\caption{The scale dependent valence quark distribution (\ref{val dis}) of the 
         lowest $(N=2)$ Fock state (long dashed and solid lines) compared with
         the GRS parametrization (dashed and dot-dashed lines). 
         As we discuss in the text to the right of the bar this parametrization
         becomes less reliable.}                   \label{plot val dis}
\end{center}
\end{figure}

We can now write down the corresponding scale dependent valence quark 
distribution. Using the asymptotic form of the pion's two-particle distribution
amplitude in expression~(\ref{pdf}) we obtain
\begin{equation}
 u_v^{(2)}(x;\mu^2)=6\,P_2(\mu^2)\,x \bar{x}, \label{val dis}
\end{equation}
which is shown in Fig.~\ref{plot val dis} at two different scales, where we
also compare with the GRS parametrization. The plot clearly shows the 
anticipated damping of the valence distribution with increasing scale.
As discussed below Eq.~(\ref{approx api}) only a few Fock states contribute
to the parton distribution at large $x$. We thus expect the valence 
distribution $u_v^{(2)}(x)$ of the lowest Fock state to approximate well the 
full valence distribution $u_v(x)$ at large $x$. 
With a fixed transverse size parameter, Fig.~\ref{plot val dis} shows
that this expectation is not fulfilled at large scales for $x$ larger than 
0.6 since $u_v(x)$ is shifted to smaller values of $x$ while $u_v^{(2)}(x)$
remains constant. We see that switching on the scale dependence of $a_\pi$ 
complies with our expectation
provided that we take into account the theoretical and experimental 
uncertainties of the analysis of the Drell-Yan process $\pi^-p\to\mu^+\mu^-X$,
from which the GRS parametrization of the valence distribution is extracted.
In particular, for $x\gsim 0.75$ the GRS valence distribution is less reliable
since in that region the parametrization of the proton structure function, 
which is used as an input in the analysis, is an extrapolation. 
Our approach provides a clear improvement compared to the results of
Ref.~\cite{JakobKrollRaulfs}, where a constant transverse size parameter has 
been used and where thus a comparison of the full and the $N=2$ parton 
distributions has been possible only at low scale. 

\begin{figure}[t]
\begin{center}
\psfig{file=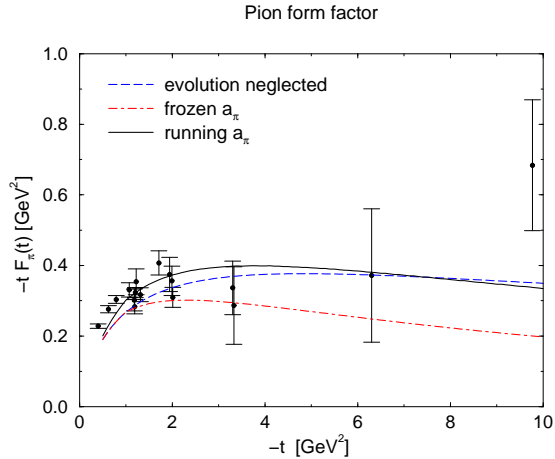,bb=45 70 560 675,width=6cm,angle=-90}
\caption{The pion form factor with constant and running $a_\pi$. The dashed 
         line shows the case of constant $a_\pi=0.86$ GeV$^{-1}$, where we have
         also fixed the scale of the GRS valence distribution at 1 GeV$^2$. 
         Data are taken from Ref.~\cite{Bebek}.} \label{plot pionff evolution}
\end{center}
\end{figure}

We will now consider the pion form factor, given by 
Eq.~(\ref{pionff}), as an application.\footnote{As can easily be shown 
numerically \cite{DFJK} the ansatz (\ref{lcwf}) of the pion's LCWF does not 
provide significant contributions to the Drell-Yan 
overlap formula of the pion form factor for $x\to 1$ and $k_\perp\to 0$ and 
so contrary to what has been claimed in \cite{Brodsky}, for instance, the LCWF
(\ref{lcwf}) does represent well soft QCD contributions.} 
Since the relevant scale in the overlap formula is given by the momentum
transfer $t=-{\bf q}_\perp^2$ with ${\bf q}_\perp^2$ being the photon 
virtuality in a frame where the plus component of the momentum transfer
vanishes, such that $\zeta=0$, it is natural to identify $\mu^2$ with $-t$. 
The result is shown in Fig. \ref{plot pionff evolution} (solid line). For 
comparison we also plot the form factor neglecting the evolution of both $q_v$
and $a_\pi$ (dashed line), and with fixed $a_\pi$  and running $q_v$ 
(dot-dashed line), respectively. The three curves 
show that the scale dependence of $a_\pi$ roughly compensates the 
evolution of the GRS valence distribution in the region 
5~GeV$^2 \le -t \le 10$~GeV$^2$. For $-t < 5$~GeV$^2$ the running $a_\pi$
even provides a slight enhancement of the theoretical prediction. 

The authors of \cite{JakobKroll,JakobKrollRaulfs} considered the contributions 
of the lowest Fock state only, without taking into account evolution effects of
the transverse size parameter. Comparing their results with the dashed line of 
Fig. \ref{plot pionff evolution}, where we have frozen the evolution scale
of both the transverse size parameter and the valence distribution, 
we see that their prediction stays below ours. This has to be expected since 
we take into account the contributions of all Fock states. 

The factorized GRS ansatz of Ref. \cite{BakulevRuskov} is essentially identical
to our expression (\ref{valence spd}), again with a scale independent 
transverse size parameter and the GRS valence distribution at a fixed scale.
Their corresponding prediction of the pion form factor  is somewhat higher than
ours. This is due to the fact that the parameter $\Lambda_0$ used in 
\cite{BakulevRuskov} corresponds to a transverse size parameter of 
$a_\pi\simeq 0.77$ GeV$^{-1}$, which is smaller than the value used in the 
present work.

The predictions of the hard contributions to $-t\, F_\pi(t)$ alone, ranging 
from 0.08 GeV$^2$ in \cite{JakobKrollRaulfs} to about 0.16 GeV$^2$ in 
\cite{Stefanis}, obviously cannot account for the experimental data. We would 
like to point out that the sum of the hard part of Ref.~\cite{JakobKrollRaulfs}
and our prediction of the soft part is in very good agreement with the new data
between 0.6 and 1.6~GeV$^2$ presented in Ref.~\cite{Volmer}. Note that in 
Ref.~\cite{Braun} strong cancellations between soft parts and hard parts of 
higher twist have been found, leaving small non-perturbative contributions.

To summarize, we have shown that the evolution equations for SPDs leads to a  
further constraint for phenomenological models of SPDs which are expressed 
through ordinary parton distributions. Starting from the generalized Drell-Yan 
formula, where the SPD for the $N$-th Fock state of the pion is written
in terms of an overlap integral of $N$-particle light-cone wave functions, and 
specializing to the case of a zero skewedness parameter the SPD equals a Fock 
state parton distribution multiplied by an exponential function. Making further
simplifications by assuming a common transverse size parameter for all Fock 
states we have obtained the full SPD, to which we have then applied the 
evolution equations. We have matched the generalized and ordinary DGLAP 
equations for the skewed and forward parton distributions, respectively, which 
has resulted in a constraint for the scale dependence of the transverse size 
parameter. This in turn has led to a scale dependent Fock state probability and
valence quark distribution of the lowest Fock state of the pion. The 
application to the soft overlap contributions to the pion form factor has shown
a slight enhancement of the theoretical prediction in the few GeV$^2$ region, 
which is in complete agreement with new data. 

Finally, we would like to remark that LCWFs of the form (\ref{lcwf}), which are
modified by an effective mass term, have also been discussed in the literature,
see, for instance \cite{BHL} and first Ref. of \cite{Stefanis}. However, our 
discussion of the 
evolution effects shows that in principle one has to take into account the 
scale dependence of the effective mass as well, which would reduce the 
influence of a mass term in the LCWF with increasing scale.\\

{\bf Acknowledgements.} 
I would like to thank Th. Feldmann and P. Kroll for stimulating discussions 
and critical comments. I have also benefited from discussions with 
A. P. Bakulev, R. Jakob, H. Huang and N. G. Stefanis. Moreover, I acknowledge 
a graduate grant of the Deutsche Forschungsgemeinschaft.


\begin{thebibliography}{99}

\bibitem{Mueller} D. M\"uller, D. Robaschik, B. Geyer, F. M. Dittes and
                  J. ${\rm Ho{\check r}ej{\check s}i}$, Fortschr. Phys. 
                  {\bf 42}, 101 (1994).

\bibitem{Ji} X. Ji, Phys. Rev. Lett. {\bf 78}, 610 (1997); 
             Phys. Rev. D {\bf 55}, 7114 (1997).

\bibitem{Radyushkin} A. V. Radyushkin, Phys. Lett. B {\bf 380}, 417
                     (1996); Phys. Rev. D {\bf 56}, 5524 (1997).

\bibitem{DFJK} M. Diehl, Th. Feldmann, R. Jakob and P. Kroll,
                Eur. Phys. J. C {\bf 8}, 409 (1999).

\bibitem{DrellYan} S. D. Drell and T.-M. Yan, Phys. Rev. Lett. {\bf 24}, 
                    181 (1970).

\bibitem{JakobKroll} R. Jakob and P. Kroll, Phys. Lett. B {\bf315}, 463 (1993).

\bibitem{BolzKroll} J. Bolz and P. Kroll, Z. Phys. A {\bf 356}, 327 (1996).

\bibitem{BHL} S. J. Brodsky, T. Huang and G. P. Lepage, 
              {\it Particles and Fields} {\bf 2}, eds. Z. Capri and 
              A.N. Kamal (Banff Summer Institute) p.\ 143 (1983).

\bibitem{Radyushkin2} A. V. Radyushkin, Phys. Rev. D {\bf 58}, 114008 (1998).

\bibitem{Afanasev} A. V. Afanasev, hep-ph/9808291.

\bibitem{BakulevRuskov} A. P. Bakulev, R. Ruskov, K. Goeke and N. G. Stefanis,
                        Phys. Rev. D {\bf 62}, 054018 (2000). 

\bibitem{Miyama} M. Miyama, S. Kumano, Comput. Phys. Commun. {\bf 94}, 185 
                 (1996). 

\bibitem{Musatov} I. V. Musatov and A. V. Radyushkin, Phys. Rev. D {\bf 56}, 
                 2713 (1997); \\ 
                 P. Kroll and M. Raulfs,  Phys. Lett. B {\bf 387}, 848, (1996).

\bibitem{CZ} V. L. Chernyak and A. R. Zhitnitsky, Nucl. Phys. B {\bf 201}, 492,
             (1982).

\bibitem{GRS} M. Gl\"uck, E. Reya, I. Schienbein, 
              Eur. Phys. J. C {\bf 10}, 313 (1999).

\bibitem{JakobKrollRaulfs} R. Jakob, P. Kroll and M. Raulfs, 
                           J. Phys. G {\bf 22}, 45 (1996). 

\bibitem{Bebek} C. J. Bebek et al. Phys. Rev. D {\bf 17}, 1693 (1978). 

\bibitem{Brodsky} S. J. Brodsky, hep-ph/9908456.

\bibitem{Stefanis} N. G. Stefanis, W. Schroers and H.-Ch. Kim, hep-ph/0005218; 
                   \\ Phys. Lett. B {\bf 449}, 299 (1999).   

\bibitem{Volmer} J. Volmer et al., nucl-ex/0010009. 
    
\bibitem{Braun} V. M. Braun, A. Khodjamirian and M. Maul, 
                Phys. Rev. D {\bf 61}, 073004 (2000).

\end{thebibliography}
\end{document}